\documentclass[amssymb,useAMS,prd,aps,amsmath,nofootinbib,superscriptaddress,
twocolumn]{revtex4-1}
\usepackage{natbib}
\usepackage{color}
\usepackage{graphicx}
\usepackage{amsmath}
\usepackage[caption=false]{subfig}
\usepackage[colorlinks=true,linkcolor=blue,citecolor=blue,urlcolor=black]{
hyperref}

\begin{document}
\title{Projections for measuring the size of the solar core with neutrino-electron scattering}
\author{Jonathan H. Davis}
\affiliation{Theoretical Particle Physics and Cosmology, Department of Physics, King's College London, London WC2R 2LS, United Kingdom
\\ {\smallskip \tt  \href{mailto:jonathan.davis@kcl.ac.uk}{jonathan.davis@kcl.ac.uk}}}
\date{\today}
\begin{abstract}
We quantify the amount of data needed in order to measure the size and position of the $^8$B neutrino production region within the solar core, for experiments looking at elastic scattering between electrons and solar neutrinos.
The directions of the electrons immediately after scattering are strongly correlated with the incident directions of the neutrinos, however this is degraded significantly by the subsequent scattering
of these electrons in the detector medium. We generate distributions of such electrons for different neutrino production profiles, and use a maximum likelihood analysis to make projections
for future experimental sensitivity. We find that with approximately $20$~years worth of data the Super Kamiokande experiment could constrain the central radius of the shell in which $^8$B neutrinos are produced to be less than $0.22$ of the total solar radius at $95\%$ confidence.
\end{abstract}
\preprint{KCL-PH-TH/2016-36}
\maketitle

\section{Introduction}
The sun is an abundant source of both electromagnetic radiation and neutrinos. It is not a point-source for either of these, with light observed from the entire sun within a region of around 0.5 degrees on the sky~\cite{Mamajek:2015jla}. Solar models predict that the neutrinos produced by
the sun originate from fusion reactions in its core~\cite{Bahcall:2004pz,Bahcall:2000nu,Antonelli:2012qu}, which should comprise approximately $20\%$ to $25\%$ of the sun's total radius $R_{\mathrm{sun}}$ according to helioseismological data~\cite{Garcia1591,Christensen-Dalsgaard1286,Basu:2009mi,Garcia:2010zt}. 
The neutrinos originate in shells of different radial positions and widths within this core, depending on the fusion reaction in which they are produced~\cite{0004-637X-765-1-14} e.g. $^8$B or pp.

However no direct measurement of the structure of this core region exists which makes explicit use of neutrino directionality.
This is because neutrinos are much more challenging to detect than electromagnetic radiation. For the most massive detectors e.g. the Super Kamiokande experiment~\cite{PhysRevD.83.052010,Abe:2016nxk,Abe:2011ts,Fukuda2003418} neutrinos are detected by looking for Cerenkov light emitted
by electrons which have been liberated from their atoms by neutrino elastic scattering. The direction of the electron immediately after scattering is strongly correlated with the incident direction of the neutrino, however its path
through the detection medium is blurred by approximately $25$ degrees due to scattering effects. Hence in practice the angular distribution of the detected electrons is only weakly correlated with the distribution of solar
neutrinos, making it very difficult to reconstruct their incident directions~\cite{Beacom:1998fj,Ando:2001zi,PhysRevD.68.093013}.

Nevertheless, for such experiments, it has been shown in the case of neutrinos from a supernova that with a large amount of statistics the incident direction of the neutrinos can still be reconstructed, and the direction of the supernova
determined to within $4^{\circ}$ or better~\cite{Beacom:1998fj,Ando:2001zi,PhysRevD.68.093013}. For supernova detection the improvement in the angular resolution of the incident neutrinos scales approximately as $N_{\nu}^{-1/2}$, where $N_{\nu}$ is the number of detected neutrino events, with the exact resolution
depending on the amount of blurring of the electron path and the size of the background. The Super Kamiokande experiment detects approximately 15 electron elastic scatter events from ${}^8$B solar neutrinos per day~\cite{Raaf_pres} and has a target mass
of approximately 22.5 kilo-tonnes of water. Hence a \emph{naive first approximation} of the smallest angular size of the sun's core detectable in principle with neutrinos is $$\theta_{\nu} \sim \frac{25^{\circ}}{\left(5475 \cdot \left[\frac{t}{\mathrm{years}}\right] 
\cdot \left[\frac{M_{\mathrm{det}}}{20 \, \mathrm{kt}}\right] \right)^{1/2}} ,$$ where $t$ is running time and $M_{\mathrm{det}}$ is the detector mass (and their product is the exposure).
This means that for an angular resolution of 0.05 degrees an experiment would need an exposure of approximately $900$ kilo-tonne years, equivalent to the Super Kamiokande experiment taking data for approximately $45$ years. In principle the Hyper Kamiokande
detector~\cite{Abe:2011ts}, which will have a fiducial volume $\sim 25$ times larger than that of Super Kamiokande~\cite{Fukuda2003418}, would only need around two years to obtain the same amount of data.

In this work we take a closer look at this scenario, to quantify how much data is required to determine the sun's core structure with $^8$B neutrinos using the angular profile and energy spectrum of the scattered electrons, and to what precision this measurement can be made in the near future.
In order to do this we numerically generate profiles of the angle of travel of the electrons, liberated from their atoms by neutrinos produced in the solar core, in a detector with respect to the centre of the sun. We then perform a two-parameter maximum likelihood analysis to determine
the statistical precision with which the central position and size of the $^8$B neutrino production shell can be measured to with a given exposure.

\section{Generating the angular distribution of scattered electrons}
\begin{figure}[t]
\centering
\includegraphics[width=0.49\textwidth]{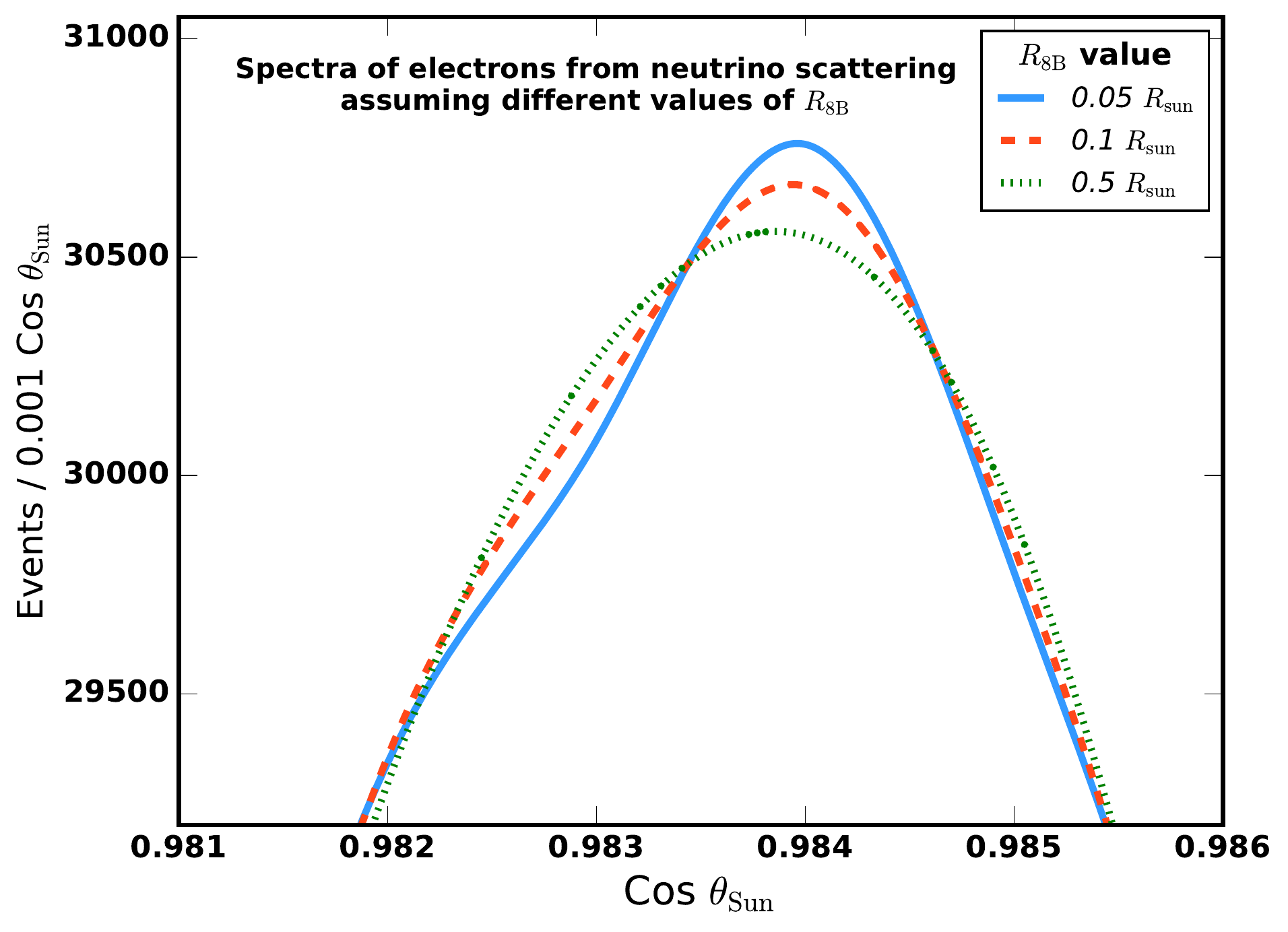}
\caption{Distribution in a neutrino detector of the angles of travel of electrons (integrated over all recoil energies above $5$~MeV) with respect to the centre of the sun, which have been liberated from their atoms due to scattering with solar neutrinos, for different values of the mean position of $^8$B neutrino production in the solar core $R_{8\mathrm{B}}$.
We take an exposure of $2000$~kton~years and assume a fixed value of $\delta_{8\mathrm{B}} = 0.05 R_{\mathrm{sun}}$.}
\label{fig:spectra}
\end{figure}

We consider the detection of $^8$B neutrinos with energies of several MeV and above via their elastic scattering with electrons bound in atoms i.e. $\nu e^- \rightarrow \nu e^-$, which is the detection mechanism used by water Cerenkov experiments such as Super Kamiokande~\cite{Abe:2011ts}.
For such experiments, which we focus on in this work, it is the electrons which are actually detected via their Cerenkov light emission, and so we need to know how the angular distribution of the neutrinos correlates with that of the detected electrons.

The distribution of the angle of scatter of these electrons $\theta_{\mathrm{sc}}$ relative to the incident
direction of the neutrinos is given by the differential cross section $\mathrm{d} \sigma / \mathrm{d} \mathrm{cos} \, \theta_{\mathrm{sc}}$. Fortunately the angular distribution of the scattered electrons is strongly correlated with the 
arrival angle of the original neutrino, especially for recoil energies above $\sim 5$~MeV~\cite{Beacom:1998fj,Ando:2001zi}, and so $\mathrm{d} \sigma / \mathrm{d} \mathrm{cos} \, \theta_{\mathrm{sc}}$ is peaked close to
$\mathrm{cos} \, \theta_{\mathrm{sc}} = 1$. The differential scattering rate is then given by the integral of this cross section over all incident neutrino energies $E_{\nu}$, multiplied by the differential flux $\Phi(E_{\nu})$ i.e.
\begin{equation}
\frac{\mathrm{d}R}{\mathrm{d} \mathrm{cos} \, \theta_{\mathrm{sc}}} = N_{\mathrm{target}} \int \mathrm{d} E_{\nu} \frac{\mathrm{d} \sigma}{\mathrm{d} \mathrm{cos} \, \theta_{\mathrm{sc}}} \Phi(E_{\nu}) ,
\label{eqn:rate}
\end{equation}
where $N_{\mathrm{target}}$ is the total number of electrons available for elastic scattering in the detector.
For the flux $\Phi(E_{\nu})$ we use the standard approximately thermal spectra from the sun~\cite{Bahcall:2004pz,Billard:2013qya}, which is dominated by $^8$B neutrinos here as we consider only electron recoil energies above 5~MeV.
\textcolor{black}{Since the differential cross section depends on the flavour composition of the incoming neutrino flux we also account for the oscillation of $\nu_e$ into $\nu_{\mu}$ and $\nu_{\tau}$ during transit from the sun to earth.
There is also the the Mikheyev-Smirnov-Wolfenstein (MSW) effect due to flavour oscillations in transit through the earth itself, leading to a $\sim 3\%$ day-night asymmetry in the detected neutrino rate~\cite{Renshaw:2013dzu}, which we neglect 
here as we consider the time-averaged rates. 
For the calculation of both the signal and background in our analysis we assume a hypothetical experiment which is identical to Super Kamiokande in both composition and the position of the site underground.}

For Super Kamiokande and similar detectors the electron scatters multiple times in the detection medium, which smears out the cone of Cerenkov light emitted by the electron,
which is used to determine the angle of its path. This smearing can be approximately modelled as a Gaussian distribution centred on the initial angle after scattering with a width which depends on energy and is between $20^{\circ}$ to $30^{\circ}$,
depending on the electron energy~\cite{Beacom:1998fj,Ando:2001zi}. However for water Cerenkov detectors such as Super Kamiokande it was shown in ref.~\cite{PhysRevD.68.093013} that the distribution of angles due to this effect is more
accurately modelled by a Landau distribution, which has a larger tail than a Gaussian. We use the latter in this work.
Hence even though equation~(\ref{eqn:rate}) is forward-peaked, the angular distribution of electrons is dominated by this multiple scattering.

Here we determine the angular distribution of the detected electrons numerically by first generating a large sample of incident neutrino directions, which depends on the size and position of the production region within the solar core.
Indeed neutrinos are expected to be produced in shells within the solar core, whose position depends on the particular fusion channel e.g. through pp or $^8$B.
\textcolor{black}{For $^8$B neutrinos, which we focus on in this work, models of the sun~\cite{0004-637X-765-1-14} predict a spherical shell with neutrino production radially distributed around a central value
close to $\mathrm{R}_{8\mathrm{B}} \approx 0.05 R_{\mathrm{sun}}$, where $R_{\mathrm{sun}}$ is the solar radius. In this work we are interested in understanding how accurately
a neutrino experiment can reconstruct this initial neutrino distribution. Hence we randomly generate the points of origin of each neutrino within a spherical shell by assuming such an initial distribution based on solar models~\cite{0004-637X-765-1-14}, but with the central value $\mathrm{R}_{8\mathrm{B}}$
and width $\delta_{8\mathrm{B}}$ set by hand independently. In this case $\delta_{8\mathrm{B}}$ is equal to twice the standard deviation of the radial distribution of neutrino production and is \emph{approximately} $0.05 R_{\mathrm{sun}}$ for the default model from ref.~\cite{0004-637X-765-1-14} . We also generate values of the recoil energy $E_r$ for each electron, which is fixed at the point of scattering by the initial neutrino energy and the size of $\theta_{\mathrm{sc}}$. }

After generating these incident directions we then calculate the scattering angles of the electrons using equation~(\ref{eqn:rate}) and account for the detector resolution using a Landau~\cite{PhysRevD.68.093013} distribution consistent with the angular distribution of electrons in 
Super Kamiokande.
We show plots of three of these distributions for different values of $R_{8\mathrm{B}}$ and with $\delta_{8\mathrm{B}} = 0.05 R_{\mathrm{sun}}$ in figure~\ref{fig:spectra} and with an exposure of $2000$~kton~years, focusing on the angles closest to the centre of the sun i.e. $\mathrm{cos} \, \theta_{\mathrm{sun}} = 1$.
For all values of $R_{8\mathrm{B}}$ the distribution of electron angles is broadly similar, however we can see from figure~\ref{fig:spectra} that a larger value of $R_{8\mathrm{B}}$ results in a distribution which is more spread out around the peak of the distribution, while
a smaller value of $R_{8\mathrm{B}}$ results in a more concentrated distribution at the peak. 

We can get an idea of the statistics required to distinguish between the $R_{8\mathrm{B}} = 0.05 R_{\mathrm{sun}}$ and $R_{8\mathrm{B}} = 0.5 R_{\mathrm{sun}}$ profiles in figure~\ref{fig:spectra} by comparing the difference between them, which is about $400$ events, to the Poisson uncertainty on the total number of events, which is around $\sqrt{100000} = 316$ near the peak of the distribution. Hence with $2000$~kton~years a neutrino experiment should
 at the very least be able to  distinguish a profile with $R_{8\mathrm{B}} = 5\%$ of the total solar radius to one with $R_{8\mathrm{B}} = 50\%$ of the solar radius. In the next section we use a maximum likelihood method to quantify the ability of neutrino 
experiments to separate electron angular profiles with different values of $R_{8\mathrm{B}}$, such as shown in figure~\ref{fig:spectra}, by exploiting the different distributions in $\mathrm{cos} \, \theta_{\mathrm{sun}}$ and $E_r$.

\section{Detectability of the solar core using neutrino-electron scattering}
\begin{figure}[t]
\centering
\includegraphics[width=0.49\textwidth]{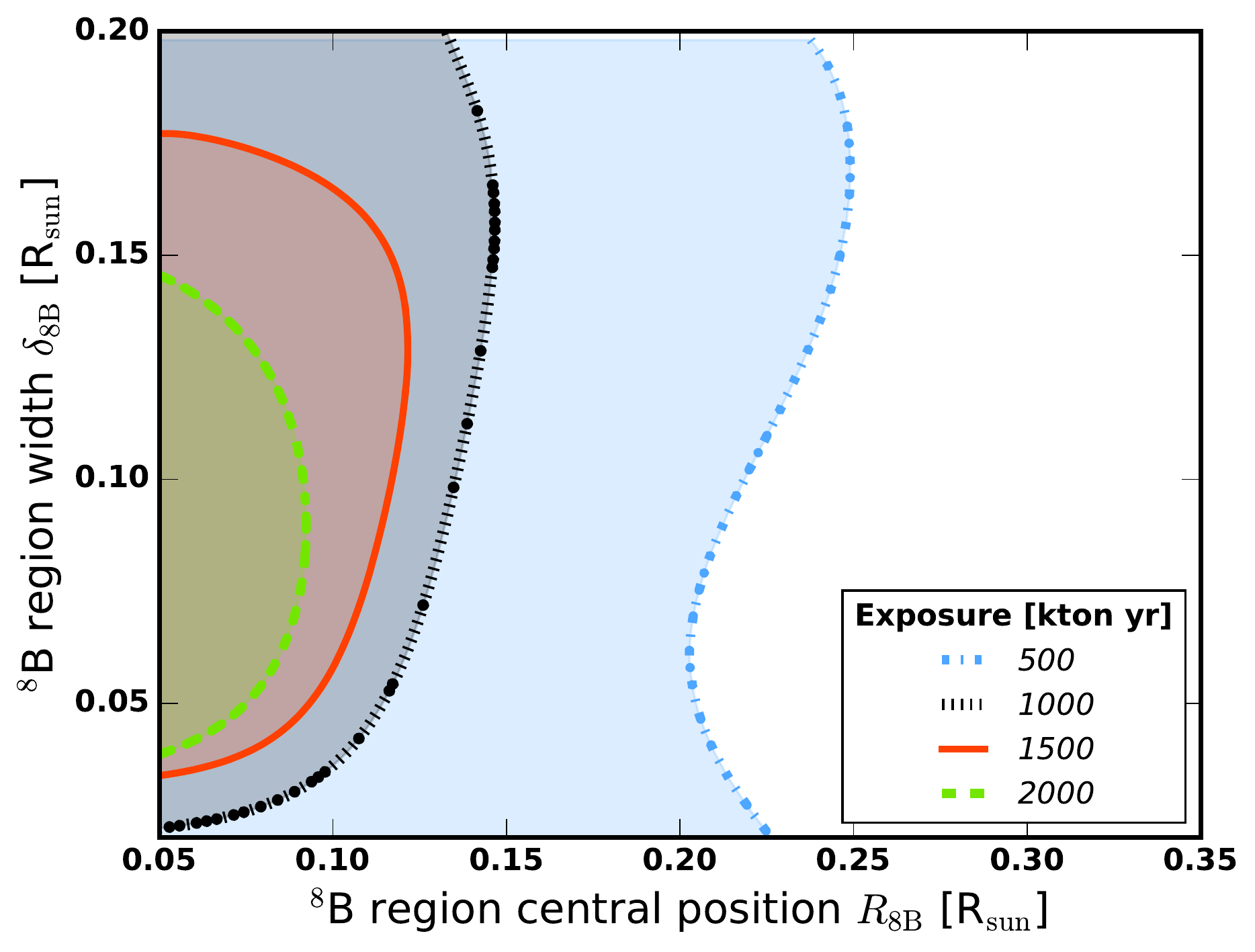}
\caption{$95\%$ confidence regions to measure the central position $R_{8\mathrm{B}}$ and width $\delta_{8\mathrm{B}}$ of the region in which $^8$B neutrinos are produced in the sun with neutrino-electron scattering, as a fraction of the total solar radius $R_{\mathrm{sun}}$. We have generated simulated data with $R_{8\mathrm{B}} = 0.05 R_{\mathrm{sun}}$ and $\delta_{8\mathrm{B}} = 0.05 R_{\mathrm{sun}}$, in line with
expectations from solar modelling~\cite{0004-637X-765-1-14,Garcia1591,Christensen-Dalsgaard1286}, and have used a maximum likelihood method to quantify how well different distributions of electrons in a neutrino detector fit to this data.}
\label{fig:res_with_exposure}
\end{figure}

In order to quantify the ability of neutrino experiments like Super Kamiokande to \textcolor{black}{measure $R_{8\mathrm{B}}$ and $\delta_{8\mathrm{B}}$} we have performed a maximum likelihood analysis, which compares simulated data to angular profiles of  electrons scattered elastically by neutrinos under different assumptions
for the size of $R_{8\mathrm{B}}$ and $\delta_{8\mathrm{B}}$. We have generated simulated data by sampling from the angular distributions of scattered electrons, according to a number of events calculated using equation~(\ref{eqn:rate}) with a cut on the electron
recoil energy of $5$~MeV. The simulated data is generated under the assumption that $R_{8\mathrm{B}} = 0.05 R_{\mathrm{sun}}$ and $\delta_{8\mathrm{B}} = 0.05 R_{\mathrm{sun}}$, similar to expectations from solar modelling~\cite{0004-637X-765-1-14,Garcia1591,Christensen-Dalsgaard1286}.
We then compare each simulated data-set to the theoretically expected angular profile of these electrons using a Poisson likelihood for each value of $R_{8\mathrm{B}}$ \textcolor{black}{and $\delta_{8\mathrm{B}}$}. We have generated distributions
from $R_{8\mathrm{B}} = 0$ up to $R_{8\mathrm{B}} = R_{\mathrm{sun}}$ in steps of $0.001 R_{\mathrm{sun}}$ and from $\delta_{8\mathrm{B}} = 0.01 R_{\mathrm{sun}}$ to $\delta_{8\mathrm{B}} = 0.2 R_{\mathrm{sun}}$ in steps of $0.01 R_{\mathrm{sun}}$. 
\textcolor{black}{We consider both the distribution in angle and recoil energy of the electrons in order to discriminate between different values of $R_{8\mathrm{B}}$ and $\delta_{8\mathrm{B}}$, since the angular spread of the recoiled electrons depends on the recoil energy, becoming
less spread out for more energetic electrons. For the recoil energy $E_r$ we consider all events with $E_r \geq 5$~MeV and use a bin-size for the analysis of $\Delta E_r = 0.5$~MeV.}
We neglect effects from time-variation of the neutrino flux~\cite{Cravens:2008aa} and all
rates here are time-averaged.

We have assumed an isotropic background rate \textcolor{black}{of approximately $5$ events per day per kilo-tonne} similar to that observed in Super Kamiokande~\cite{PhysRevD.83.052010}, since it has been shown that a large background can inhibit the ability to accurately determine the incident direction
of the neutrinos in the case of supernovae~\cite{Beacom:1998fj,Ando:2001zi,PhysRevD.68.093013}. 
We also consider the spectral shape of this background in recoil energy based on results from the Super Kamiokande collaboration~\cite{PhysRevD.83.052010,Abe:2016nxk,Abe:2011ts,Fukuda2003418}.
Indeed if the rate of background events is comparable to the signal rate then the resolution of the detector can be significantly degraded. \textcolor{black}{In our case, based on the results from Super Kamiokande-III~\cite{PhysRevD.83.052010}, the overall background rate is about six times the size of the signal rate. However it is smaller than the signal rate within the range of scattering angles
closest to the centre of the sun.}

\textcolor{black}{Shown in figure~\ref{fig:res_with_exposure} is the $95\%$ confidence region for \textcolor{black}{measuring $R_{8\mathrm{B}}$ and $\delta_{8\mathrm{B}}$} as a function of exposure. 
With $500$~kton years of exposure the Super Kamiokande experiment could measure $R_{8\mathrm{B}} \lesssim 0.22 R_{\mathrm{sun}}$ at $95\%$ confidence and with $2000$~kton years this constraint improves to a level of $R_{8\mathrm{B}} \lesssim 0.09 R_{\mathrm{sun}}$, while 
also placing a bound on $\delta_{8\mathrm{B}}$ within the region $0.04 R_{\mathrm{sun}} \lesssim \delta_{8\mathrm{B}} \lesssim 0.15 R_{\mathrm{sun}}$.
The former exposure requires a running time of just over 20~years for Super Kamiokande, while the latter exposure would require a running time of more than $80$~years.}

\textcolor{black}{In principle the Hyper Kamiokande experiment, which will have
a fiducial mass in excess of $500$~kilo-tonnes~\cite{Fukuda2003418,Kearns:2013lea}, could achieve this 2000~kton~year exposure in a much more realistic time-frame. However Hyper Kamiokande is not simply a larger version of Super Kamiokande, and so it may for example 
have a larger background (due to a potentially shallower site) or a different directional resolution, making a direct comparison difficult. An experiment at a different site would also receive a different flavour composition of neutrinos
due to the MSW effect, which would alter the time-dependence of the flux over the day/night cycle.
Hence it is more appropriate to consider instead a hypothetical experiment, which we call ``SK500'',
which is an experiment identical to Super Kamiokande, and with identical conditions, but with a fiducial mass of 500~kilo-tonnes.
For SK500 the latter exposure would be achievable in only $4$~years. }

\textcolor{black}{These results could likely be improved by the Super Kamiokande collaboration in a dedicated study optimised for understanding the solar core structure, using the advanced analysis techniques introduced in their recent phase IV run~\cite{Abe:2016nxk}. Specifically in the phase IV run the low-energy
threshold for electron recoils is reduced to 3.5~MeV and both the angular resolution and signal-to-background ratio are improved using a statistic called the Multiple-Scattering Goodness (MSG). 
We are not able to perform an analysis using the MSG statistic, however we can estimate its effect by observing that it results in a signal-to-background ratio which is approximately twice as large for larger values of the MSG statistic and for angles closest to the sun.
With a background reduced by a factor of two relative to the solar neutrino rate we find that the constraint on $R_{8\mathrm{B}}$ improves to $R_{8\mathrm{B}} \lesssim 0.075 R_{\mathrm{sun}}$ at $95\%$ confidence with $2000$~kton~years of data. A dedicated analysis performed by the
Super Kamiokande collaboration could expect at least this increase in sensitivity over our results, and potentially more if they can improve the effective angular resolution with the MSG statistic.}

\section{Conclusion}
Experiments looking for solar neutrinos, such as Super Kamiokande~\cite{PhysRevD.83.052010,Abe:2016nxk}, can measure the elastic scattering between neutrinos and electrons to obtain directional information on the neutrino source~\cite{Beacom:1998fj,Ando:2001zi,PhysRevD.68.093013}. 
However this is made significantly more difficult since the resulting electrons
scatter in the detector, leading to their measured angles of travel being only weakly correlated with those of the incident neutrinos.

In this work we have made quantitative projections for neutrino detectors with poor directional resolution, focusing on those based on detection of Cerenkov light from electrons such as Super Kamiokande.
We have determined the amount of data required to study the region within the sun in the sky from which the solar neutrinos originate, using the angular profile and energy spectrum of electrons scattered
elastically by solar neutrinos. 

We have focused here on $^8$B neutrinos, since these will produce the most events in water Cerenkov detectors above their thresholds of around $5$~MeV. 
Solar models~\cite{0004-637X-765-1-14,Garcia1591,Christensen-Dalsgaard1286} predict that they originate from a spherical shell within the solar core with central position $R_{8\mathrm{B}} \approx 0.05 R_{\mathrm{sun}}$ and width $\delta_{8\mathrm{B}} \approx 0.05 R_{\mathrm{sun}}$.
Hence we constrain both $R_{8\mathrm{B}}$ and $\delta_{8\mathrm{B}}$ using a maximum likelihood analysis.
This provides complimentary information to helioseismological measurements of the solar core~\cite{Garcia1591,Christensen-Dalsgaard1286,Basu:2009mi}, and could help in solving the discrepancies between models of the sun and experimental data~\cite{Serenelli:2016nms,Lopes:2001ra,Vincent:2014jia,PenaGaray:2008qe}.
It could also be combined with information on neutrino oscillations~\cite{Lopes:2013sba}.

As shown in figure~\ref{fig:spectra} we have calculated the resulting angular profiles of electrons from $^8$B solar neutrinos by using a Landau distribution, which models the angular resolution of 
water Cerenkov detectors such as Super Kamiokande~\cite{PhysRevD.68.093013}. Using these distributions, at different recoil energies, we performed a maximum likelihood analysis to quantify how well $R_{8\mathrm{B}}$ and $\delta_{8\mathrm{B}}$ can be measured with a given experimental exposure, by fitting different profiles of scattered electrons as a function of scattering angle and recoil energy to simulated data.
The result is shown in figure~\ref{fig:res_with_exposure}.
\textcolor{black}{After scanning over a wide range of $R_{8\mathrm{B}}$ and $\delta_{8\mathrm{B}}$ values we find that with approximately $20$~years worth of data the Super Kamiokande experiment could place a limit of $R_{8\mathrm{B}} \lesssim 0.22 R_{\mathrm{sun}}$ at $95\%$ confidence.
Looking at future experiments, with approximately $4$ years of data-taking an \textcolor{black}{experiment identical to Super Kamiokande but with a target mass of $500$~kilo-tonnes} could place a limit of $R_{8\mathrm{B}} \lesssim 0.09 R_{\mathrm{sun}}$ and $0.04 R_{\mathrm{sun}} \lesssim \delta_{8\mathrm{B}} \lesssim 0.15 R_{\mathrm{sun}}$ at $95\%$ confidence.}

However it remains to be seen whether this is achievable for a real experiment, such as Super Kamiokande, as we have assumed that the only uncertainties on determining $R_{8\mathrm{B}}$ are due to the statistical effects brought on by
the scattering of the electrons, assuming a somewhat idealised form for the distribution of these electrons. \textcolor{black}{Fortunately much progress has been made by the Super Kamiokande collaboration in reducing their systematic uncertainties.
For example as shown in Table IV of ref.~\cite{Abe:2016nxk} there is a systematic uncertainty in the
modelling of the angular resolution of Super Kamiokande, which for the most recent phase IV run manifests as less than a $1\%$ discrepancy between data and Monte Carlo. This is smaller than the variation in angular
resolution due to different values of $E_r$ within the $0.5$~MeV bins used in our analysis, which is at least one or two degrees, and so is unlikely to alter our results significantly.} 

\textcolor{black}{The Hyper Kamiokande experiment is a potential candidate for improving on Super Kamiokande in this regard, since it will have at least $25$ times the target mass~\cite{Abe:2011ts,Fukuda2003418}. However it is not simply a larger version of Super Kamiokande and it would also need to have a signal-to-background
ratio and directional resolution at least as good as Super Kamiokande in order for our estimates to be accurate.}
Even so we have shown that if a detector has poor directional resolution it can still place useful constraints on the structure of the solar 
core using directional information, with the detection of a large number of neutrino events.  
Hence a high-resolution image of the sun using neutrinos could open important new avenues for solar physics and neutrino physics.

\section*{Acknowledgements}
The research leading to these results has received funding from the European Research Council through the project DARKHORIZONS under the European Union's Horizon 2020 program (ERC Grant Agreement no.648680).

\end{document}